\documentstyle[twocolumn,pra,aps,epsfig]{revtex}
\begin{document}
\flushbottom
\draft
\title{Gravity-Induced Wannier-Stark Ladder in an Optical Lattice }
\author{J. Zapata$^{1,2}$, A. M. Guzm\'an$^{2}$, M. G. Moore$^1$, and P. Meystre$^1$}
\address{$^1$Optical Sciences Center, University of Arizona, Tucson, Arizona
85721\\
$^2$Departamento de F\'\i sica, Universidad Nacional de Colombia, Bogot\'a.}
\maketitle
\begin{abstract}
We discuss the dynamics of ultracold atoms in an optical potential accelerated
by gravity. The positions and widths of the Wannier-Stark ladder of resonances
are obtained as metastable states. The metastable Wannier-Bloch states
oscillate in a single band with the Bloch period. The width of the resonance
gives the rate transition to the continuum.
\end{abstract}
\section{INTRODUCTION}
Optical lattices are a cornerstone of much of atom optics. They
have been successfully used to form matter-wave diffraction
gratings, were at the heart of the discovery of subrecoil cooling,
and were manipulated to study atomic Bloch oscillations \cite{rb},
and dynamic instabilities and chaos.\cite{rr} They are also used
as an outcoupling mechanism in a mode-locked atom laser \cite{rk},
and are central to several quantum computing schemes using neutral
atoms.\cite{rj} As atom optics rapidly moves toward applications,
frequently listed possibilities include gravimeters and gravity
gradiometers. Gravity is also an important and natural way to
accelerate atoms, so as to coherently reduce their de Broglie
wavelength for nanofabrication applications. Hence, there is
considerable interest in studying the combined effects of an
optical lattice and gravitation on the dynamics of ultracold
atoms.

The study of particles in periodic potentials has of course been a
major tenet of condensed matter physics, and it will therefore
come as no great surprise that our analysis relies heavily on the
use of its tools. In particular, delocalized Bloch states and band
structure analyses are known to be ideal for the study of
electrons in crystals. For deep potentials, though, the use of
localized Wannier states is often preferable and indeed, has
proven useful e.g. in the study of the dipole-dipole interaction
in optical lattices.\cite{ra} When a linear potential is
superimposed to the underlying periodic structure, the situation
is less clear: the localized Wannier states become delocalized,
and there is some debate as to which representation is then most
useful. One possibility is to use a so-called Wannier-Bloch states
representation, with an emphasis on the delocalized character of
the atomic wave functions. Alternatively, the Wannier-Stark
approach emphasizes the localized aspect of the wave functions. A
third alternative consists in treating the periodic potential as a
perturbation on top of the gravitational field. Clearly, each
approach has its pros and cons, depending upon the depth of the
lattice compared to the gravitational potential. For deep
lattices, though, the Wannier-Stark analysis seems preferable. In
this approach, the states of the atoms form a sequence of
resonances separated by equal energy intervals referred to as
Wannier-Stark ladders.\cite{rw,rav,ri} Such ladders were recently
observed in accelerated optical lattices. \cite{rt1,rt2}

In this paper we calculate the Wannier-Stark ladders induced by gravity on
atoms in optical lattices, using a theoretical approach recently developed
in Ref. \cite{r10} and based on a Floquet representation of the Wannier states.
This technique, which leads to a well-defined eigenvalue problem for the
determination of the Wannier-Stark ladder, was developed for the more general
situation of time-periodic potentials, but is also perfectly adapted to
time-independent potentials.

Section II introduces our physical model, and briefly reviews
the definitions of the Bloch and Wannier states, which are the
eigenstates of the system in the absence of a gravitational field. Section III
outlines the Floquet approach to the determination of the
Wannier-Stark ladder, discussing the onset of a ladder of complex ``resonances''
associated with metastable localized states of the atoms. Selected numerical
results are presented in section IV. Finally section V is a summary and outlook.

\section{PHYSICAL MODEL}

We consider the situation of ultracold atoms of mass $M$ trapped
in the optical potential resulting from two far-off resonant and
counterpropagating optical fields of frequency $\omega_L$ and wave
number $k_L = 2\pi/\lambda_L$. The atoms are assumed to be
two-level atoms, and we neglect all complications resulting from
the possible existence of multiple ground states and the
associated optical pumping. After adiabatic elimination of their
upper electronic states, the atoms are therefore described as
scalar particles subject to the periodic potential
\begin{equation}
{\hat V}({\hat z})=\frac{V_0}{2} \cos(2 k_{L}{\hat z})
\label{a1}
\end{equation}
where
\begin{equation}
V_0=\frac{\hbar \Omega^2}{4(\omega_L-\omega_A)}
\end{equation}
where $\Omega$ is the Rabi frequency of the atomic transition of frequency
$\omega_A$ under consideration.
In the presence of gravity, the atomic Hamiltonian of the system is given by
\begin{equation}
\hat{H}=\hat{H}_0 + Mg{\hat z},
\label{a2}
\end{equation}
where $Mg$ is the force of gravity and
\begin{equation}
\hat{H}_0=\frac{\hat{p}^2}{2M}+{\hat V}({\hat z}).
\label{a3}
\end{equation}
We consider a one-dimensional situation where the lasers generating the
optical lattice are along the $z$-axis, so that $[{\hat z},{\hat p}] = i\hbar$.

The band structure associated with the Hamiltonian $\hat{H}_0$ is
of course well-known. It is displayed in Fig. 1 for reference. It
shows the energy-wave number dispersion relation $E_n(k)$ in the
first Brillouin zone \cite{rss}, $n$ being the band index. The
associated energy eigenstates of $\hat{H}_0$ are the familiar
Bloch states
\begin{equation}
\hat{H}_0\phi_{n,k}(z) = E_n(k)\phi_{n,k}(z).
\label{a4}
\end{equation}
They are conveniently expressed as
\begin{equation}
\phi_{n,k}(z)=\exp(ikz)\chi_{n,k}(z) ,
\label{a99}
\end{equation}
where the functions $\chi_{n,k}(z)$ are periodic in $z$. They
satisfy the eigenvalue equation
\begin{equation}
\hat{H}_0^{(k)}\chi_{n,k}(z)=E_n(k)\chi_{n,k}(z),
\label{a5}
\end{equation}
the Hamiltonian $\hat{H}_0^{(k)}$ being given by
\begin{equation}
\hat{H}_0^{(k)}=\frac{(\hat{p}+\hbar k)^2}{2M}+V(z).
\label{a51}
\end{equation}

From Eq. (\ref{a5}), the evolution of the Bloch wave functions $\chi_{n,k}(z)$
is simply
\begin{equation}
\chi_{n,k}(z,t)= \exp[-iE_n(k)t/\hbar] \chi_{n,k}(z,0).
\end{equation}
It is furthermore possible to take advantage of the periodicity of
the optical potential to expand $\chi_{n,k}(z)$ on a lattice basis
$\{|\ell \rangle \}$, $\ell$ integer, as the Fourier series
\begin{equation}
\chi_{n,k}(z) = \sqrt{\frac{k_L}{\pi}}\sum_{\ell} c_\ell^{(n,k)}
\exp(2i\ell k_L z),
\label{a6}
\end{equation}
where we have used the coordinate representation of $\{ | \ell
\rangle \}$,
\begin{equation}
\langle z|\ell \rangle=\sqrt{\frac{k_L}{\pi}}\exp(2i\ell k_L z),
\end{equation}
and we recognize the primitive reciprocal vector $b = 2k_L$ in the
exponent.

In prevision of our subsequent discussion, we introduce at this point the
characteristic time
\begin{equation}
T_B = \frac{2\hbar k_L}{Mg} ,
\label{tb}
\end{equation}
which will later on be identified as the ``Bloch period,'' as well as
the associated ``Bloch phase''
\begin{equation}
\Theta_n(k) = E_n(k)T_B/\hbar .
\label{theta}
\end{equation}
Since the energies $E_n(k)$ always appear in the form of a phase factor in
the evolution of the system, it is in many cases convenient to plot them
in a reduced phase interval. This representation will be useful in identifying
the resonances appearing in the presence of a gravitational potential.

We conclude this section by recalling that the Bloch states are completely
delocalized, with the atomic wave functions
extending over the whole lattice. For deep lattices, it is oftentimes
more convenient to use instead localized Wannier states. They are constructed
as linear superpositions of the Bloch states $\phi_{n,k}(z)$ as
\begin{equation}
\varphi_{n,m}(z) =  \int_{-k_L}^{k_L}\exp(-iR_mk) \phi_{n,k}(z)dk,
\label{a7}
\end{equation}
where $R_m$ is the position of the $m$-th lattice site,
\begin{equation}
R_m = m \pi/k_L ,
\end{equation}
$m$ is an integer, and we recognize that $\pi/k_L$ is just the primitive
lattice period $a$ of the optical potential. As is well known, the Wannier
function $\varphi_{n,m}(z)$ describes an atomic wave packet localized at
the $m$-th site of the optical lattice.

\section{WANNIER-BLOCH STATES AND WANNIER-STARK STATES}

We now turn to a discussion of the impact of gravity on the atomic dynamics
in the lattice. We proceed by eliminating the gravitational contribution
from the Hamiltonian (\ref{a2}) via the unitary transformation
\begin{equation}
\psi (z,t)=\exp[-iMg{\hat z}t/ \hbar ]\tilde{\psi}(z,t) .
\label{b1}
\end{equation}
The evolution of $\tilde{\psi}(z,t)$ is governed by the {\em time-dependent}
transformed Hamiltonian
\begin{equation}
\tilde{\cal H}(t)= \frac{(\hat{p}-Mgt)^2}{2M}+{\hat V}({\hat z}),
\label{b2}
\end{equation}
which is nothing but the Hamiltonian $\hat{H}_0$ with momenta
shifted by $Mgt$. From this observation, it follows that its time
evolution is governed by the equation
\begin{equation}
\tilde{\cal H}(t+T)=\exp(iMgT{\hat z}/ \hbar)\tilde{\cal H}(t)
\exp(-iMgT{\hat z}/ \hbar ).
\label{b3}
\end{equation}
Eq. (\ref{b3}) provides an immediate understanding of Bloch
oscillations: After a time $T_B=2\hbar k_L/Mg$, the momentum shift
appearing in the Hamiltonian (\ref{b3}) is precisely equal to
$2k_L$, the primitive reciprocal lattice vector. An atom starting
at the edge of the first Brillouin zone is therefore uniformly
accelerated into the next Brillouin zone, or equivalently
undergoes an umklapp process in the first zone. In optics terms,
this means that the Bragg scattering condition is fulfilled, with
scattering of the atom from the state of momentum $k_L$ to $-k_L$.
This justifies identifying $T_B$ as the Bloch period as we have
done in section II.

With this intuition at hand, we introduce the Wannier-Bloch states
$\psi_{n,k}(z)$ as those states which are stroboscopic eigenstates of the
time-dependent Hamiltonian (\ref{b3}) at successive intervals of $T_B$, that is,
that satisfy the eigenvalue equation
\begin{equation}
\hat{\cal U}(T_B)\psi_{n,k}(z)=\exp\left [-\frac{i}{\hbar}E_n(k)T_B\right ]\psi_{n,k}(z).
\label{b5}
\end{equation}
with
\begin{equation}
\hat{\cal U}(T_B)= \exp \left (-\frac{i}{\hbar} \int_{0}^{T_B}\hat{H} dt \right ).
\label{b4}
\end{equation}

Despite the fact that we use the same notation $E_n(k)$ for
simplicity, it is important to realize that the quantities
appearing in Eq. (\ref{b5}) are {\em not} the same as the band
energies of section II. Indeed, due to the fact that the
Hamiltonian in the presence of gravity is time-dependent, the
$E_n(k)$ are only quasi-energies, or ``resonances.'' We will see
shortly that they are complex, their imaginary part leading to a
finite lifetime, and form a series of ladders of equidistant
levels.

The eigenstates $\psi_{n,k}(z)$ present the considerable advantage
that the only momentum involved at the end of their stroboscopic
evolution is the primitive reciprocal lattice vector $2k_L$. That
is, $\hat{\cal U}(T_B)$ commutes with the momentum shift operator
$\exp[-i\pi \hat{p}/\hbar k_L]$, and hence it is still possible to
use a Bloch-like quasi-momentum representation to stroboscopically
diagonalize the problem.  Rewriting then, in analogy with Eq.
(\ref{a99}), the Wannier-Bloch functions as
\begin{equation}
\psi_{n,k}(z)=\exp(ikz)\chi_{n,k}(z) ,
\end{equation}
where $\chi_{n,k}(z)= \chi_{n,k}(z+\pi/k_L)$, leads to the Bloch-like
eigenvalue equation \cite{r10}
\begin{equation}
\hat{\cal U}^{(k)}\chi_{n,k}(z)=
\exp\left [-\frac{i}{\hbar}E_n(k)T_B \right ]\chi_{n,k}(z),
\label{b7}
\end{equation}
where the operator $\hat{\cal U}^{(k)}$ is given by
\begin{equation}
\hat{\cal U}^{(k)}=e^{-2ik_Lz}\tilde{\cal U}^{(k)}(T_B)
\label{b77}
\end{equation}
and
\begin{eqnarray}
& &\tilde{\cal U}^{(k)}(T_B)= \nonumber \\
& &{\cal T} \exp \left [-\frac{i}{\hbar}\int_{0}^{T_B}\left (
\frac{(\hat{p}+\hbar k-Mgt)^2}{2M}+V(z)\right ) dt \right ],
\label{b8}
\end{eqnarray}
where ${\cal T}$ stands for time-ordered.

From the analysis so far, it would appear that we need to solve
the eigenvalue problem (\ref{b7}) for a large number of initial
times between 0 and $T_B$ to obtain the full system evolution.
This, however, is not the case, as the quasi-energy spectrum
$E_n(k)$ is completely degenerate for the time-independent
potential considered here, $E_n(k) = E_n$. This can be seen by
considering the evolution of a Wannier-Bloch state for a fraction
$T=T_B/s$ of the Bloch period, where $s$ is an integer. In that
interval the operator
\begin{equation}
\hat{\cal U}_T=e^{-2ik_Lz/s}\tilde{\cal U}(T),
\label{b9}
\end{equation}
with
\begin{equation}
\hat{\cal U}_{T}^{s}=\hat{\cal U} ,
\label{b54}
\end{equation}
transforms the Wannier-Bloch state $\psi_{n,k}(z)$ into another
Wannier-Bloch state of quasi-momentum $k' = k - 2k_L/s$,
\begin{equation}
\hat{\cal U}_T \psi_{n,k}(z)=\psi_{n,k^\prime}(z).
\label{b55}
\end{equation}
Combining Eqs. (\ref{b54}) and (\ref{b55}) yields readily
\begin{eqnarray}
&&\hat{\cal U}\psi_{n,k'}(z) = \hat{\cal U}\hat{\cal U}_T \psi_{n,k}(z)
= \hat{\cal U}_T \hat{\cal U}\psi_{n,k}(z)
\nonumber \\
&&= \hat{\cal U}_T \exp \left [-\frac{i}{\hbar} E_n(k)T_B \right ]
\psi_{n,k}(z)\nonumber \\
&& = \exp\left [-\frac{i}{\hbar} E_n(k)T_B \right ] \psi_{n,k'}(z)
\end{eqnarray}
where we have used Eq. (\ref{b54}) to commute $\hat{\cal U}$ and
$\hat{\cal U}_T.$ This completes the proof that for the
time-independent potential ${\hat V}(z)$ at hand, the quasi-energy
spectrum is degenerate.

As a result, the continuous time evolution of the Wannier-Bloch
states is simply given by
\begin{equation}
\psi_{n,k}(z,t)=\exp(-iE_n t/\hbar)\psi_{n,k-Mgt/\hbar}(z,0).
\label{b99}
\end{equation}

The calculation of the quasi-energies $E_n$ follows the procedure
discussed in Ref. \cite{r10}. It consists in finding the
metastable Wannier-Bloch states by solving the eigenvalue problem
(\ref{b7}) assuming a truncation of the momentum space. The
representation of the operator $\hat{\bf{\cal U}}^{(k)}$ in this
truncated basis is nonunitary, and for this reason the
``energies'' $E_n$ are complex,
\begin{equation}
E_n \rightarrow E_n - i\Gamma_n/2 .
\end{equation}
Their real part give the positions of the ``resonances'', while
their imaginary part give their widths, or lifetimes. This is to
be contrasted to the gravitation-free situation, where the
eigenvalue problem is tridiagonal in momentum space, and the
eigenvalues are then real. Eq. (\ref{b8}) shows clearly how the
acceleration of gravity removes the discrete, tridiagonal nature
of the problem and makes it continuous in momentum space.

We have previously discussed how the Bloch functions approach puts
the emphasis on delocalized states. Clearly, this same
characteristic also holds for the Wannier-Bloch states discussed
so far. As in the gravitation-free situation, it is sometimes
useful to consider states of a localized nature instead. In
complete analogy with the Wannier functions, we therefore
introduce the Wannier-Stark states as
\begin{equation}
\Psi_{n,m}(z)=\int_{-k_L}^{k_L} \exp(-iR_mk) \psi_{n,k}(z)dk,
\label{b11}
\end{equation}
where as we recall $R_m = m\pi/k_L$ is the position of the $m$-th lattice
site. Taking into account the temporal evolution (\ref{b99}) of the Wannier-Bloch
states gives the continuous time evolution of the Wannier-Stark
wave functions $\Psi_{n,m}(z,t)$,
\begin{equation}
\Psi_{n,m}(z,t)=\exp[-\frac{i}{\hbar}(E_n+ m \pi Mg/k_L)t]\Psi_{n,m}(z).
\label{b12}
\end{equation}
This shows that the Wannier-Stark states constitute a ladder of
quasi-stationary functions characterized by sets of discrete levels of energies
\begin{equation}
E_{n,m} = E_n+m \pi Mg/k_L,
\label{ladder}
\end{equation}
of equal spacings $\pi Mg/k_L$. These energy levels are illustrated in
Fig. 2.

\section{RESONANCES}

In this section, we present selected results from the numerical
study of gravitationally-induced Wannier-Stark resonances for the
case of shallow optical lattices. The calculations proceed by
solving the eigenvalue equation
\begin{eqnarray}
&&\hat{\bf{\cal W}}^{(k)}\chi_{n,k}(z)=\nonumber \\
&&\exp\left(-\frac{\Gamma_n(k)}{2\hbar}T_B \right )\exp\left (-
\frac{i}{\hbar}E_n(k)T_B\right )\chi_{n,k}(z),
\label{b13}
\end{eqnarray}
where $\hat{\bf{\cal W}}^{(k)}$ is the operator (\ref{b77}), truncated in
momentum space as previously discussed.

To diagonalize the matrices $\hat{\bf{\cal W}}^{(k)}(\ell^\prime,\ell)$,
we first evaluate the operator $\tilde{\cal U}^{(k)}(T_B)$ using the formula
\begin{equation}
\tilde{\cal U}^{(k)}(T_B)={\cal T}\prod_{j=0}^{N-1}\,
\exp \left [-\frac{i}{\hbar}\int_{t_j}^{t_{j+1}}\tilde {\cal H}^{(k)}(t) dt \right ],
\label{c1}
\end{equation}
where
\begin{equation}
\tilde {\cal H}^{(k)}(t)=\frac{(\hat{p}+\hbar k-Mgt)^2}{2M}+V(z),
\label{c2}
\end{equation}
and $t_{j+1}-t_j=\Delta t=T_B/N$ ($N>>1$). The exponential factors
in (\ref{c1}) are approximated by a Dyson series truncated at
second order,
\begin{eqnarray}
\label{c3}
& &\exp \left [-\frac{i}{\hbar}\int_{t_j}^{t_{j+1}}\tilde {\cal H}^{(k)}(t) dt
\right ]\approx 1+ (-\frac{i}{\hbar})\int_{t_j}^{t_{j+1}}\tilde {\cal H}^{(k)}(t) dt
\nonumber \\
& &+(-\frac{i}{\hbar})^{2}\int_{t_j}^{t_{j+1}}\tilde {\cal H}^{(k)}(t) dt
\int_{t_j}^{t}\tilde {\cal H}^{(k)}(t') dt'.
\end{eqnarray}
The analytic evaluation of the matrix elements of the Dyson series
(\ref{c3}) is a straightforward procedure. We just give the result
of the second term:
\begin{eqnarray}
& & \langle \ell^\prime|\int_{t_j}^{t_{j+1}}\tilde{\cal H}^{(k)}(t) dt|\ell \rangle =
\frac{E_RT_B}{N}\left [ 2\ell+\kappa-\frac{2j+1}{N}\right ]^2 \delta_{\ell^\prime,\ell}
\nonumber \\
& &+\frac{E_RT_B}{3N^{3}}\delta_{\ell^\prime,\ell}+ \frac{V_0T_B}{4N}
\left ( \delta_{\ell^\prime,\ell +1}+\delta_{\ell^\prime,\ell-1}\right ),
\label{c4}
\end{eqnarray}
where $E_R=\hbar^{2}k_{L}^{2}/2M$ is the recoil energy,
$\kappa=k/k_L$ and we have used the definition (\ref{tb}) of the
Bloch period. For fixed $\kappa$, Eq. (\ref{c3}) gives the
elements of a symmetric matrix whose truncation causes no problem.
We proceed by first truncating it, and by an iterative process of
successive time-ordered intervals we then evaluate the evolution
operator $\tilde{\cal U}^{(k)}(T_B)$ using Eq. (\ref{c1}). The
matrix elements of the operator $e^{-2ik_Lz}$ are given by
\begin{equation}
\langle \ell^\prime|e^{-2ik_Lz}|\ell \rangle=\delta_{\ell^\prime,\ell -1},
\label{c5}
\end{equation}
whose truncation generates a nonunitary operator. Taking the
dimension of the matrices to be $2R+1$, we have found that for
sufficiently large $R$, typically, 20 to 40, the position of the
resonances becomes independent of the dimension of the truncated
momentum space.

The simulations were carried out for the case of $^{87}Rb$ atoms
\cite {rk} trapped in an optical lattice generated by a far-off
resonance optical standing wave of wavelength $\lambda=850nm$. The
depth of the lattice was taken to be $V_0=-2.1 E_R$, where $E_R
\approx 3.177$ kHz is the recoil energy. This choice of parameters
corresponds to a level separation of $\Delta E_{ws} = \pi
Mg/k_L=0.2854485 E_R$ in the Wannier-Stark ladder, see Eq.
(\ref{ladder}).

Fig. 3 shows two resonances, folded in the interval $-\pi
<\Theta_n(k)< \pi$ where $\Theta_n(k)=E_{n}(k)T_{B}/\hbar$ (modulo
$2\pi$), see Eq. (\ref{theta}). They are degenerate in $k$, as
discussed in section III. (Recall that only degenerate solutions
are associated with resonances.) The real part of these
quasienergies are $E_0=0.10771571 E_R$ and $E_1=1.024990 E_R$,
respectively, corresponding to the Bloch phases $\Theta_0=2.37112$
(solid line) and $\Theta_1=-2.57103$ (dotted line). The widths of
the resonances, given by the imaginary part of the degenerate
eigenvalues of Eq. (\ref{b13}), are shown in Fig. 4. For the
relatively shallow lattice considered here, the highest resonance
is probably too wide, with $\Gamma_1/2=7.182\times 10^{-2}E_R$
(dotted line), to be of practical use. In contrast, the width of
the most stable resonance, $\Gamma_0/2=2.17378\times 10^{-4}E_R$
(solid line), is three orders of magnitude narrower than the
Wannier-Stark ladder separation $\Delta E_{ws}$. As further
discussed in the following section, such narrow lines might be of
considerable interest in the development of atom optical gravity
gradiometers.

Despite the long lifetime of the metastable Wannier-Stark states, the
fact that they are finite indicates that atoms initially localized
in a lattice site eventually tunnel to neighboring wells. This is shown
in Fig. 5, which plots on a logarithmic scale the absolute squared expansion
coefficients $|c_{\ell}^{(n,k=0)}| ^2$ of the Bloch states, see Eq. (\ref{a6}),
for the lowest band $n=0$. The solid line correspond to the
gravity-free situation, while the dotted line correspond to gravity.
When the gravity force is applied the metastable states tunnel through the
optical potential hills in the downward direction (negative $z$-axis), so
the contribution of negative $2\ell k_L$ vectors increases. Indeed,
there is a general shift of all the coefficients $c_{\ell}^{(n,k)}$
due to the action of the gravity field.

We have also evaluated the position and width of the resonances as
a function of the well depth $V_0$ of the optical lattice. Figs. 6
and 7 show the position and width of the lowest resonance, $E_0$
and $\Gamma_0/2$ respectively, in units of $E_R$. As we can see in
these plots, the position $E_0$ goes up as the well depth
decreases, while, the width $\Gamma_0/2$ decreases very rapidly
for highest potential depths. Finally, Figure 8 is a plot of the
energy gap $E_{gap}$ between the lowest resonance and continuum
states.

\section{Summary and conclusions}

For a number of potential applications of atom optics, such as e.g.
nanolithography and atom holography, it is desirable to accelerate the
atoms so as to produce coherent beams of shorter de Broglie wavelength. This
can be achieved either via gravity, or in lattices with ramped detuning between
the two counterpropagating fields, as already mentioned. Since eventually, the
atoms will then escape the trap, matter-wave acceleration is also of much
relevance in the study of output couplers for atom lasers, as was realized
for instance the mode-locked laser of Ref. \cite{rk}. In addition, atom optics
with accelerated beams offers novel approaches to interferometry. For instance,
using optical fields to hold atoms against gravity may lead to the
realization of accelerometers of unprecedented sensitivity.

These new developments require a detailed understanding of the combined
effects of light and gravity on the motion of ultracold atoms.
Using a recently developed Floquet-type analysis \cite{r10} that removes many
of the question marks previously attached to the computation of the Wannier-Stark
ladder, we have shown that a detailed quantitative calculation of the position and
width of these levels can now easily be obtained numerically. We illustrated how
deep optical lattices lead to the existence of families of
long-lived, equidistant quasi-energy levels that should be of much
interest in gravitation gradiometers and accelerometers. As the wells become
shallower, however, the number of quasi-bound levels decreases, and their
lifetime shortens.

We note that the applications we have in mind are likely to
eventually require the use of atomic samples, most likely
Bose-Einstein condensates, for which atom-atom interactions and
the associated nonlinear phase shifts can no longer be ignored.
This is likely to influence the Wannier-Stark spectra in
interesting ways, that we plan on studying in the future.

\acknowledgments{This work is supported in part by Office of Naval
Research Contract No. 14-91-J1205, National Science Foundation Grant
PHY-9801099, the Army Research Office and the Joint Services Optics Program.
JZ thanks COLCIENCIAS, Universidad del Atl\'antico and Universidad del Norte
for financial support. Discussions with J. Heurich and E. M. Wright are
greatfully acknowledged.}

--------------------------------------------------------------------
\newpage

\begin{figure}
\includegraphics[width=1.0\columnwidth]{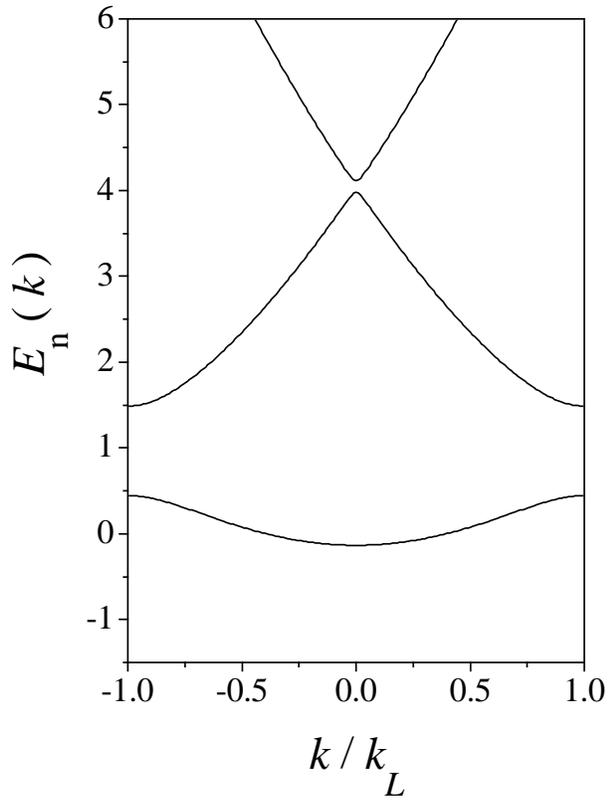}
\caption{Energy bands $E_n(k)$ (in units of $E_R$) of an optical lattice as function
of the quasimomentum $k$ in the first Brilouin zone $[-k_L,k_L]$ for $V_0 = -2.1E_R$.}
\end{figure}

\begin{figure}
\includegraphics[width=1.0\columnwidth]{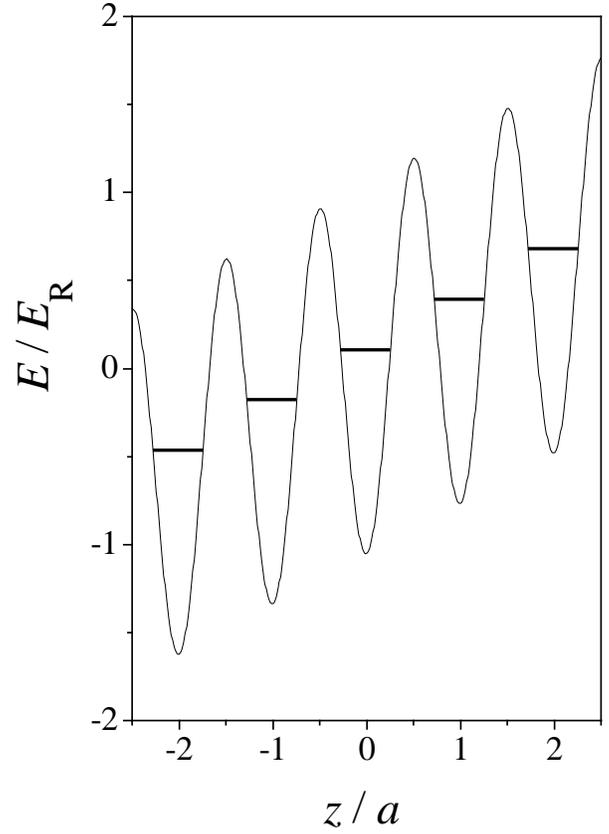}
\caption{Tilted periodic potential and Wannier-Stark ladders, in
units of the recoil energy $E_R$, as functions of $z$ for $V_0 =
-2.1 E_R$.}
\end{figure}

\begin{figure}
\includegraphics[width=1.0\columnwidth]{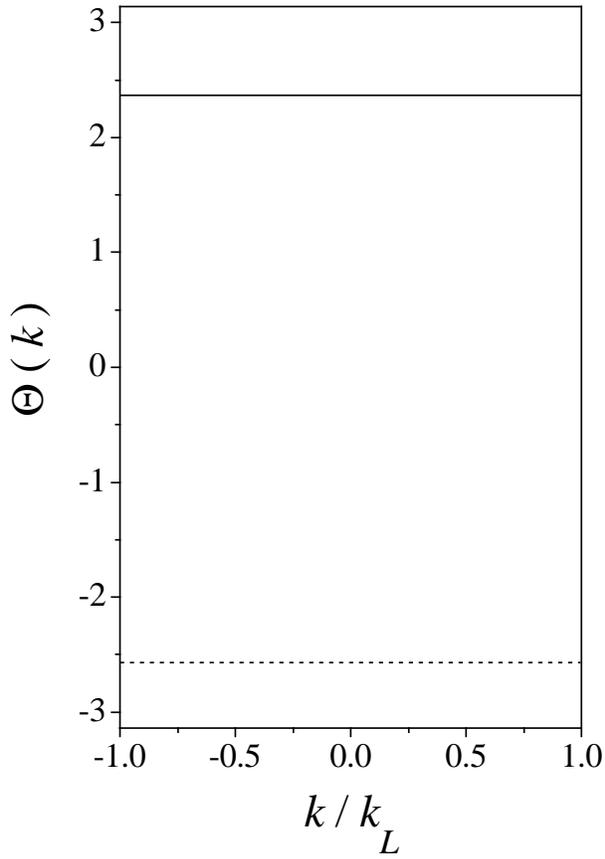}
\caption{Numerical results for resonances folded in the interval
$-\pi < \Theta(k)< \pi$ with $\Theta(k)=E_{n}(k)T_{B}/\hbar$
(modulo $2\pi$) for $V_0=-2.1E_R$. $\Theta_0=2.37112$ (solid line)
and $\Theta_1=-2.57103$ (dotted line). They correspond to energies
$E_0=0.10771571 E_R$ and $E_1=1.024990 E_R$, respectively. }
\end{figure}

\begin{figure}
\includegraphics[width=1.0\columnwidth]{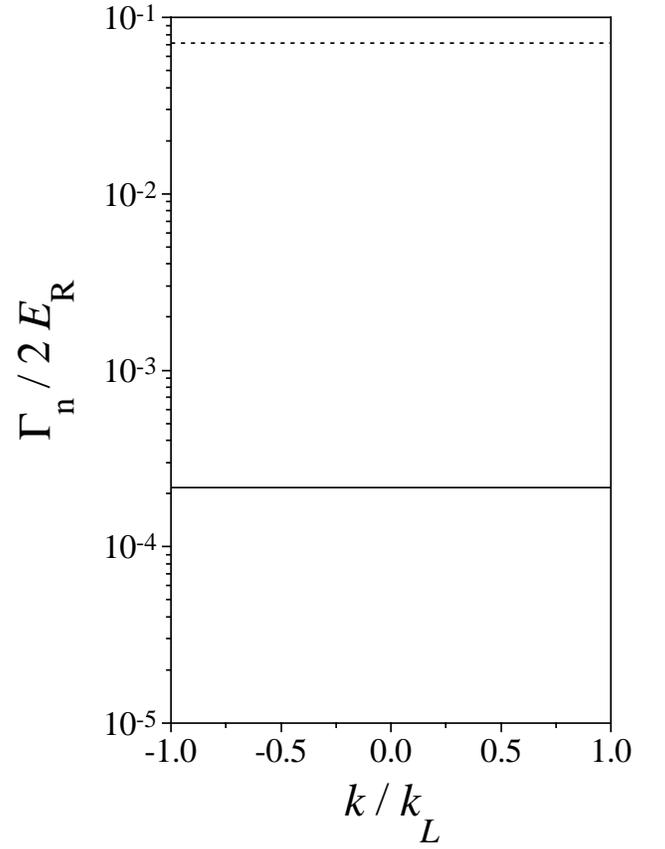}
\caption{Numerical results for the width of the resonances $\Gamma_{n}(k)/2$
(in units of $E_R$). $\Gamma_0/2=2.17378\times 10^{-4}E_R$ (solid line)
and $\Gamma_1/2=7.182\times 10^{-2}E_R$ (dotted line).}
\end{figure}

\begin{figure}
\includegraphics[width=1.0\columnwidth]{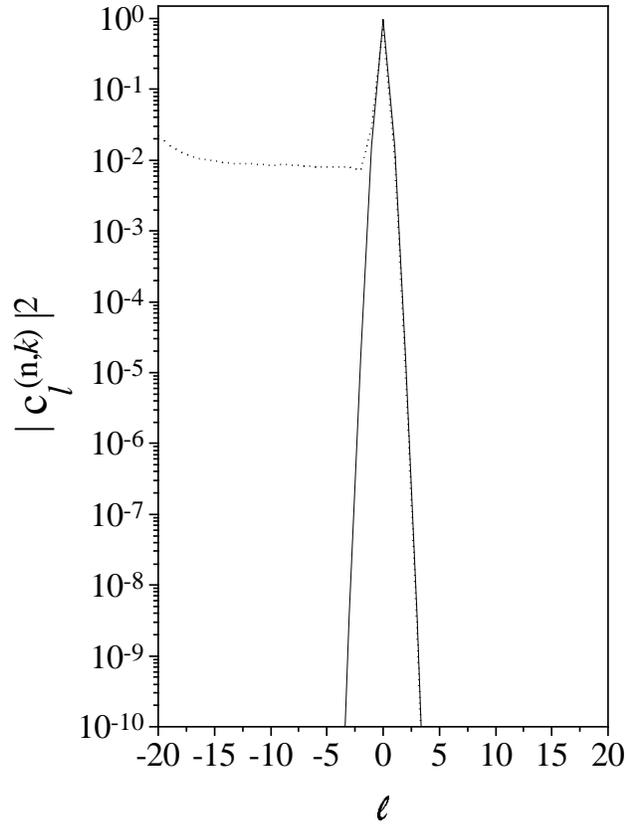}
\caption{Absolute squared coefficients $|c_{\ell}^{(n,k)}| ^2$ on a
logarithmic scale for band $n=0$ and quasimomentum $k=0$ in the field-free
case (solid line) and gravity case(dotted line).}
\end{figure}

\begin{figure}
\includegraphics[width=1.0\columnwidth]{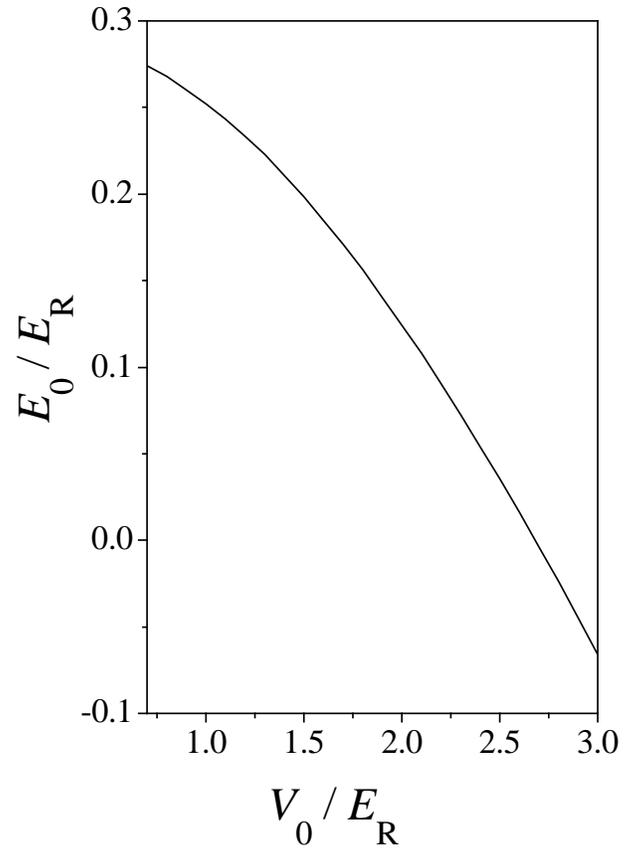}
\caption{Lowest resonance $E_0$ (in units of $E_R$) as function of the potential depth
$V_0$.}
\end{figure}

\begin{figure}
\includegraphics[width=1.0\columnwidth]{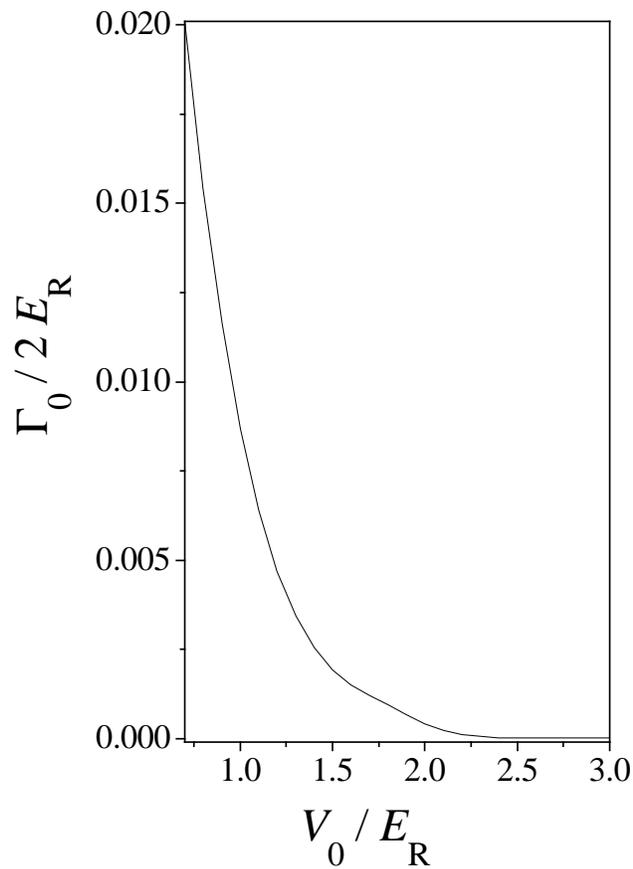}
\caption{Width $\Gamma_0/2$ (in units of $E_R$) as function of the potential depth
$V_0$.}
\end{figure}

\begin{figure}
\includegraphics[width=1.0\columnwidth]{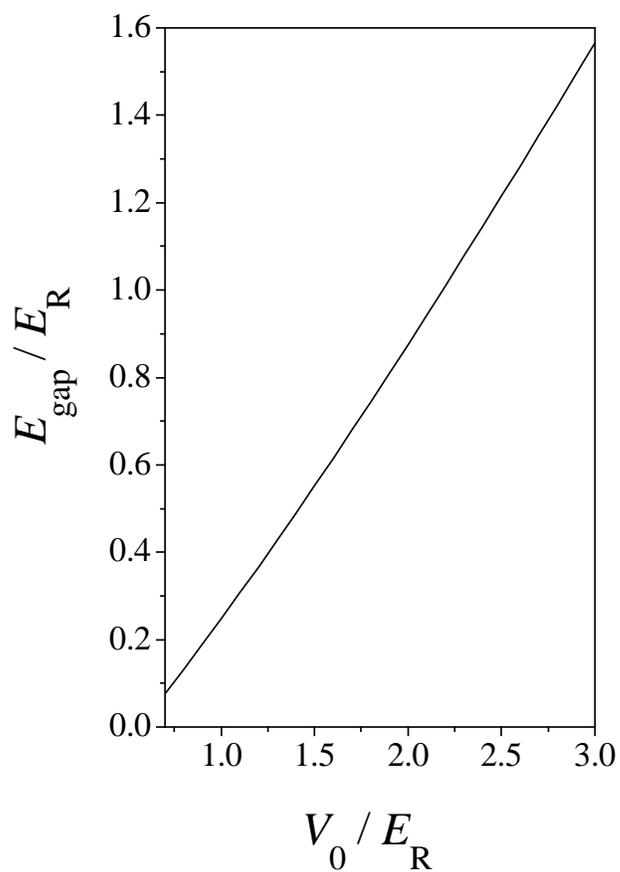}
\caption{Energy gap $E_{gap}$ (in units of $E_R$) between the lowest resonance and
continuum states as function of the potential depth
$V_0$.}
\end{figure}

\end{document}